# Using Camouflaged Cyber Simulations as a Model to Ensure Validity in Cybersecurity Experimentation

Carrie Gardner
*Software Engineering Institute
Carnegie Mellon University;
School of Computing and
Information Sciences
University of Pittsburgh*
Pittsburgh, USA
carriegardner428@gmail.com

Abby Waliga
*School of Computing and
Information and Graduate
School of Public and
International Affairs
University of Pittsburgh*
Pittsburgh, USA
abby.waliga@pitt.edu

David Thaw
*School of Law, School of
Computing and Information, and
Graduate School of Public and
International Affairs
University of Pittsburgh*
Pittsburgh, USA
dbthaw@gmail.com

Sarah Churchman
*School of Computing and
Information
University of Pittsburgh*
Pittsburgh, USA
smc161@pitt.edu

*Abstract*—Experimental research methods describe standards to safeguard scientific integrity and reputability. These methods have been extensively integrated into traditional scientific disciplines and studied in the philosophy of science. The field of cybersecurity is just beginning to develop preliminary research standards and modeling practices. As such, the science of cybersecurity routinely fails to meet empirical research criteria, such as internal validity, external validity, and construct validity. These standards of experimentation enable the development of metrics, create assurance of experimental soundness, and aid in the generalizability of results. To facilitate such empirical experimentation in cybersecurity, we propose the adaptation of camouflaged cyber simulations as an approach for cybersecurity research. This research tool supports this mechanistic method of experimentation and aids in the construction of general cybersecurity research best practices.

*Keywords—Cybersecurity, Cybersecurity Experimentation, Research Methods, Modeling, Cyber Simulation, Honeypot*

## I. Introduction

Cybersecurity currently suffers from an array of empirical experimentation problems due to a lack of research methodology standards [1], [2]. Other scientific disciplines have formal procedures to address experimentation problems such as selection bias, recency, and sampling bias [3]. Researchers in cybersecurity have increasingly discussed the importance of standardized methodology for experimentation, but standardized methodologies have been slow to emerge [4]. This lack of standardization then perpetuates reactive rather than proactive science. This paper examines the utility of using cyber simulations as a mechanism for empirical studies in cybersecurity. This paper examines: (1) the status of experimentation in cybersecurity, (2) the need for a robust method for scientific experimentation, and (3) how cyber simulations can be used as an effective empirical approach.

## II. Current Methodological Gaps in Cybersecurity Experimentation

One of the fundamental elements of any discipline is the method by which it establishes and evaluates propositions as reliable and sound [5]. Scientific disciplines usually require rigorous standards for experimentation to preserve research integrity and ensure findings are objective and reliable [6]. Professions that depend on trust in the scientific and research and development process, such as medicine or architecture, require by virtue methodological integrity checks and robust experimental best practices in order to assure operational safety and security. Dependable research and science guides technological development and best practices. Cybersecurity, like other other applied fields, relies on this chain of trust to ensure fidelity.

Many subfields in computer science have developed formal methods of proof both for theoretical constructs and for empirical approaches which evaluate observed findings. Yet, the broader field of cybersecurity lacks such a unifying methodology or structure [7]. An initial response to this absence might be to assume that the methods unique to each subfield will work together in the aggregate to create a single methodology of cybersecurity. This approach, however, wrongly assumes that the constituent methods will be compatible and will not create problematic overlap in their composition. Such an approach invokes the fallacy of composition [8], as some norms of a given set (e.g., cryptography) do not directly infer patterns of its superset (e.g., cybersecurity).

Several recent papers have illuminated this gap in cybersecurity experimentation and have proposed methods or

Research funded in part by the University of Pittsburgh Office of the Senior Vice Chancellor for Research, the School of Computing and Information, and the School of Law.

standards for normalization. Hatleback and Spring describe the need for a mechanistic approach to cybersecurity experimentation [9] and propose using a Cyber Kill-Chain [10] as one such model. Maxion [11] discusses the plague of confounding variables with loose experimental procedures and how cybersecurity experimentation must strive for transparent reporting to increase the fidelity of scientific findings. Rossow et al. report [12] on a survey of malware research studies published in academic venues and discuss the methodological shortcomings of much of the work, specifically focusing on the lack of internal and external validity in the surveyed studies. This literature sample corroborates our claim that the field of cybersecurity needs empirical standardization and rigor.

Cybersecurity needs a standard methodology for experimentation and research evaluation. Three (non-exhaustive) core tenets of scientific inquiry include internal validity, external validity, and construct validity [13]. These criteria are well established tenets of rigorous experimentation and safeguard scientific integrity. Internal validity asks the question of whether or not the evidence demonstrates a causal relationship or merely only a correlational one [14]. External validity asks whether evidence generalizes to other contexts and examples [14]. Construct validity asks whether the evidence presented supports the conclusions drawn from that evidence [13]. Any methodology (or science of) cybersecurity must rigorously apply these standards of empirical experimentation to mechanisms of scientific research so that the conclusions drawn are both accurate and applicable across the discipline. Accordingly, an experimental approach must be configurable to control and regulate the testing environment to ensure these principles are satisfied.

### III. THE NEED FOR A ROBUST MODEL FOR SCIENTIFIC EXPERIMENTATION

Existing empirical research in cybersecurity has been limited by various constraints on collecting data, such as the need to maintain security of production systems, the cost of deploying sensor networks, and privacy concerns [15]. These constraints exemplify just some of the limitations of experimenting on live production environments that drive researchers to explore other experimental methods. Common alternative measures include anecdotal examination (e.g., post-incident analysis) and static dataset analysis (e.g., network traffic or malware analysis). These approaches, however, risk generating variation in the empirical dependability of the experiment as they can inherently introduce induce experimental design flaws and bias. An example flaw would be generalizing results for risk management practices from anecdotal post-incident analysis. This methodology fails to account for Maxion's concern regarding confounding variables, and general principles of external validity.

Overcoming these constraints is critical not only as a matter of scientific integrity, but also because these experimental findings form the foundation upon which professional fields develop best practices used to perform critical tasks. This gap in cybersecurity experimentation can be remedied through more robust experimental models that can be configured to emulate various cybersecurity scenarios and environments [16]. In theory, such models could be environments which simulate live production systems deployed in live public network environments (i.e., "in the wild") to enable rigorous scientific testing. Such environments would be built employing a dynamically configurable design. This configurability allows for dynamic emulation of various environments in order properly to simulate specific architectures and systems for each given experiment. These configurations thus qualify as traditional experimental controls. In practice, these models are a substantially advanced re-invention of an old cybersecurity technology known as a honeypot, or (collectively) honeynets.

A honeypot is an intentionally vulnerable system designed to gather information on attacks that occur against it. Rudimentary ("low interaction") honeypots generally gather limited data on a single attack vector with no or little dynamic system activity [17]. More advanced ("high interaction") honeypots may introduce some automated system activity or examine multiple attack vectors [18]. Honeynets are networks of honeypots designed to emulate production environments for the purpose of studying attacker techniques without risking production systems or exposing the design of the experiment [19]. Historically, honeynets were limited as a model for experimentation due to resource constraints [20]. Once identified or "fingerprinted" by attackers, the empirical validity of a given set of honeynets would be weakened if not completely undermined [19]. Further, even some of the more sophisticated examples of high-interaction honeynets still failed to capture meaningful, comprehensive pictures of system activity because of their inherent limitation in emulation capabilities.

A high-fidelity simulation that is indistinguishable from a production environment can collect this type of comprehensive data. The failure to distinguish these types of data collection undermines the internal validity with a data collection tool by in a manner which can exclude alternate hypotheses regarding indicators and mechanisms of compromise [19]. This shortcoming was due in part to the limitations and cost of implementations based on static hardware and software. Today, advances in virtualization and automation technologies, along with their widespread adaptation into organizations of all sizes, have enabled the implementation of robust cyber simulations as virtualization and is no longer considered an inherent red flag to attackers. These advanced cyber simulations utilize virtualization and automation technologies for rapid deployment, configuration, and sophisticated emulation in order to avoid detection and fingerprinting.

Cyber simulations with proper implementation can be combined with advances in camouflage and data capture to fill the empirical gap of current cybersecurity scientific methods. Such systems are capable of full-scope data collection which addresses empirical validity issues by capturing the complete picture of system and attacker activity. At a high level, knowing the nature and scale of attack types would help an organization improve its cybersecurity defensive effectiveness, and how to deploy simulations to gather that information.

These simulations can improve the scope of information by collecting a full picture of system activity in real-time rather than, for example, being limited to network traffic. For example, consider the case of a hypothetical web-based travel business which conducts customer credit card transactions. Attackers would likely assume sensitive consumer financial data would exist within that organization, making the simulated business an attractive target. In this simulation example, all system state changes would be deterministic as the design only emulates human behavior, and thus critical files, logs, and other data could manageably be tracked by simulation monitoring functions. Any deviations would necessarily be the result of "unplanned" changes, and could be flagged for analysis.

Detection of these changes can be accomplished through a variety of methods, for example in the case of files on the local system by employing hashing algorithms on a periodic schedule. Hash changes not anticipated by the simulation would trigger flags, which in turn would indicate that the underlying data and any other simulation observation data should be passed to the analytics system for analysis. Because the system is designed as a completely virtualized small business network, there is a "perfect whitelist" of traffic and activity that is considered the system baseline. By capturing network flow, system and application logs, and virtual machine "snapshots", any aberrant activity can be tracked from start to finish as a deviation from that baseline. This creates a very high signal-to-noise ratio for anomaly detection and, thus, a strong likelihood of capturing the full picture of attacker system interaction. Thus, subsequent analysis can utilize full-scope and timely attack information allowing identification of both the vectors and vehicles used in a specific attack whereas a non-controlled environment introduces a litany of other variables that could alter and obfuscate conclusions due to scenario-specific externalities.

These simulations can also address the problems of recency and selection bias which currently limit post-incident analysis. For example, when ransomware is introduced into the wild, its initial impact can be have a devastating, crippling effect to personal and business systems. As outlined in [21] the vulnerability lifecycle goes through several phases: innovation, commercialization, and social gain. Innovation reflects the discovery of a vulnerability [21]. Commercialization occurs after the initial use of a vulnerability but before said vulnerability has become well-known [21]. The final stage of the lifecycle occurs after the vulnerability has been addressed and properly patched; however, individuals or companies who do not update their security can still be vulnerable [21]. An old vulnerability can still be crippling if security is not kept up to date.

Once a vulnerability is identified, malware scanners can be updated with the ransomware's unique signature to prevent a machine from being susceptible to the attack. However, a change to the ransomware's code or function will thereby change the ransomware's signature potentially allowing it to escape detection by malware scanners. Enterprising attackers may attempt to add to or modify the code to account for a new attack vector, to suit their intended target, or merely to avoid signature detection. Regardless of motivation, any such change would interfere with signature-based detection. Until the malware definitions are updated to include the variant, users can be susceptible to attack. There variants can still cause harm even after a subsequent patch has been released to the users. Attackers continue to find new methods of leveraging existing technology to propagate an attack across a network. Some of these methods target legitimate means to administrate systems [22], thereby handicapping administrators and reducing the means to which automate an enterprise.

Identifying older variants from post-incident analysis thus may not provide sufficiently recent data to protect against present and future threats. Furthermore, the use of "live" production systems to observe attacker activity requires risking the exposure of those systems to compromise. Properly camouflaged cyber simulations, by contrast, would not expose production systems to this risk and could allow malware and other threats to fulfill their full lifecycle allowing both for real-time and full-scope analysis of the attack process and vulnerability vectors.

Thus, properly-designed cyber simulations can address empirical validity concerns present with other existing methods of cybersecurity data collection. Since cyber simulations, by design, capture attacker data while allowing for errors generated by the user and local system, the empirical utility of collected data has a highly favorable signal-to-noise ratio. These controls help ensure internal validity by reducing the likelihood that user activity, local system error, and other non-attacker variables induce confounding variables like the ones Maxion describes. The camouflage functions of honeypots, which are designed to emulate production systems, bolster the external validity of data captured with properly-camouflaged cyber simulations. Effective camouflage ensures that captured attack data describes the actions of adversaries intent on compromising live production systems, thus making such findings more reliable and generalizable [23].

Additionally, these configurable cyber simulations allow researchers to control the risk profile of systems dynamically. This flexibility allows controlled, focused testing of different vulnerabilities (variables) in attack surfaces. Variations in attack surfaces thus become the experimental, or independent variables, which combines with the controls for internal and external validity described above to allow researchers to directly test potential relationships between attack surface variation and likelihood of system compromise. This empirical design mechanism helps safeguard the construct validity of the experimental process, leading to more reliable recommendations for cybersecurity practices [24].

The process of creating iterative designs of the generic test environment to address specific cyber security hypotheses is akin to standard object-oriented programming processes. The generic test environment is the class and each instance of the generic test environment is a separate hypothesis with its own environmental parameters. These class parameters describe high-level features of an operational environment that shape normal behavior. We consider this set of features the independent variables of the simulation design. This design process is cyclically refined as data is collected, the hypothesis is modified, and the test instance is updated. (See Fig. 1)

GENERIC TEST ENVIRONMENT

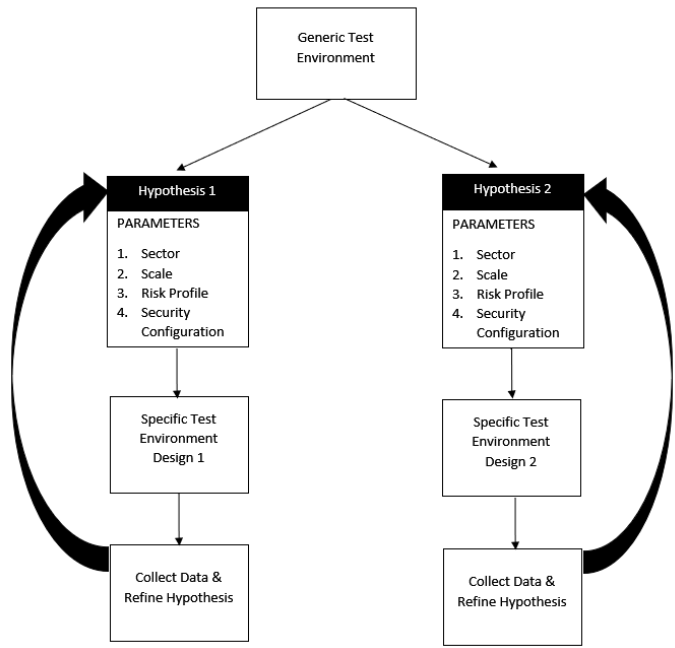

Fig. 1. *Iterative design of testing envrionments*

IV. TESTING THE MECHANISM: THE CHAMELEON PROJECT

This paper introduces an approach to use camouflaged simulations as a cybersecurity data collection tool and describes preliminary findings. In the CyREN Laboratory at the University of Pittsburgh, the Chameleon Project configures virtualized cyber simulations that emulate live production systems to collect comprehensive data, including both full-scope system activity and network activity analysis.

Data capture methods are implemented through logging sensors on hosts, network devices, and application services. This method of full-scope data collection uniquely distinguishes this cyber simulation platform from other cyber simulation technologies [18], [23], [24]. All the information is processed, analyzed, and archived into an open-source data analytics platform. The analysis platform enables researchers to identify critical events in the cyber simulation environment from various vantage points, aiding in incident detection and analysis.

The camouflage tactics of the Chameleon Project serve both to conceal the research intentions from attackers and to preserve external validity. It is imperative that these tactics remain hidden, and thus, only a high level overview is provided.

The Chameleon model seeks to emulate production environments in order to present potential adversaries with environments nearly indistinguishable from live production targets. In contrast to traditional honeypots, which usually emulated single services [18]. The Chameleon cyber simulation emulates entire environments (hosts, network activity, and services) to simulate real world organizations, such as businesses or governments, comprehensively and realistically. This project aims to capture current or advanced malicious tactics by deceiving attackers into believing emulated systems are valuable targets. Reliance on emulation techniques also supports the external validity criterion, as it acts as a control to ensure the project's findings can be applied to real-world systems, many of which now employ virtualization in cloud-based services.

Over the course of the first year and a half, Chameleon has successfully attracted attacks appropriate to the size, scope, and complexity of its attack surface. Chameleon has successfully captured data regarding the character and methods of these attacks. While further analysis of this data is reserved for future work, initial results strongly indicate the viability of cyber simulations as an effective model for empirical cybersecurity research.

V. FUTURE WORK

The analysis presented in this paper and the efficacy of the initial Chameleon tests suggest that investment in controlled experiments of broader scale is warranted using virtualized and camouflaged cyber simulations with environments like Chameleon. Future work will investigate the development of robust, scalable methods of deploying camouflaged cyber simulation systems as well as automating system activity and data collection. Such work will develop and test specific hypotheses regarding the efficacy of extant and proposed cybersecurity methods, practices, and tools. One specific example currently under study is the comparative efficacy of consumer-grade versus more advanced network perimeter defense tools. Chameleon cyber simulation models are well suited to conduct this empirical experiment by emulating various production environments and testing security configurations dynamically. This type of research illustrates the advanced and unique modeling capabilities of Chameleon. Making scalable, custom modeling for small businesses is a future priority of the lab.

When designing these simulations, reviewing similar business models and existing sites that have been breached in the past is imperative to ensure realism and develop a storefront that presents a realistic target to would-be attackers. Simulations must include physical business locations that cannot be invalidated by reconnaissance and valid communication mediums (i.e., telephone numbers and emails) automated by scripts to operate like a legitimate business. To implement realistic personas, profiles will contain what an employee is expected to do as well as how he/she may actually work. Each employee will have their own personal interests, which will be based on a set of unique personalized standard applications. For instance, an employee may use iTunes or another multimedia player and the Chrome browser instead of Safari. Due to different employee working hours, the traffic generated will vary according to each employee and their profile.

## VI. Conclusion

While previous cyber simulation technologies attempted to enable reputable scientific modeling, the failure to implement proper controls and configurations prevented these instruments from meeting the scientific standards of experimentation. Chameleon cyber simulations overcome these shortcomings and achieves external validity through effective camouflage practices. Other cybersecurity collection methods currently fall short of meeting empirical soundness.

Consequently, the Chameleon project demonstrates the plausibility of properly camouflaged cyber simulations as an empirical model for scientific experimentation. This research enables the progress of a more methods-based approach to experimentation in the cybersecurity field. It reintroduces cyber simulations as a practical instrument for data collection and investigates proper modeling techniques for cybersecurity research to meet empirical validity standards. In future work, this research will aid in informing cybersecurity risk management and decision making through evidence-based, customized modeling which is not presently possible due to the lack of features in the current cyber simulation models.

## Acknowledgment

The authors thank Abdulaziz Alarfaj, Gerry Bella Jr., Tyler Brooks, Jeremy Chavin, Ryan Craig, Adeline Giritharan, Harold Gosney, Jaclyn Joyce, Chinmay Lele, Hobart Richey, Stephen Skoch, and James Weslow for their contributions to this research.